\documentclass{llncs}
\usepackage{amsmath}
\usepackage{booktabs} 
\usepackage{graphicx}
\usepackage[all]{nowidow}
\usepackage[utf8]{inputenc}
\usepackage{tikz}
\usetikzlibrary{er,positioning,bayesnet}
\usepackage{multicol}
\usepackage{algpseudocode,algorithm,algorithmicx}
\begin{document}
\title{Improved security solutions for DDoS mitigation in 5G Multi-access Edge Computing}
\author{Marian Gușatu \and
Ruxandra F. Olimid}
\institute{Department of Computer Science, University of Bucharest, Romania
\email{marian.gusatu@unibuc.ro}, \email{ruxandra.olimid@fmi.unibuc.ro}}
\maketitle          
\begin{abstract}
Multi-access Edge Computing (MEC) is a 5G-enabling solution that aims to bring cloud-computing capabilities closer to the end-users. This paper focuses on mitigation techniques against Distributed Denial-of-Service (DDoS) attacks in the context of 5G MEC, providing solutions that involve the virtualized environment and the management entities from the MEC architecture. The proposed solutions are an extension of the study carried out in \cite{dynamic-management} and aim to reduce the risk of affecting legitimate traffic in the context of DDoS attacks. Our work supports the idea of using a network flow collector that sends the data to an anomaly detection system based on artificial intelligence techniques and, as an improvement over the previous work, it contributes to redirecting detected anomalies for isolation to a separate virtual machine. This virtual machine uses deep packet inspection tools to analyze the traffic and provides services until the final verdict. We decrease the risk of compromising the virtual machine that provides services to legitimate users by isolating the suspicious traffic. The management entities of the MEC architecture allow to re-instantiate or reconfigure the virtual machines. Hence, if the machine inspecting the isolated traffic crashes because of an attack, the damaged machine can be restored while the services provided to legitimate users are not affected.   


\end{abstract}
\section{Introduction}

A continuously increasing number of users in the online environment and stricter performance requirements (e.g., low latency) demand major changes in the new generations of networks. These lead to the necessity of new technologies that allow a high availability of services and security techniques that combine various areas such as computer science, telecom, and others. An important role in capitalizing on the promises in the 5th generation of mobile networks (5G) is played by the Multi-access Edge Computing (MEC), which aims to bring the cloud closer to the end-users. MEC has applicability in various fields, including the Internet of Things (IoT), autonomous cars, and virtual reality. The European Telecommunications Standards Institute (ETSI) introduces this concept in 2014 under the name Mobile Edge Computing \cite{mec2014}. Subsequently, in 2017, the name is changed to Multi-access Edge Computing, highlighting the acceptance of a variety of access technologies and thus making use of additional advantages~\cite{mec2018}.


Due to its novelty and impact, all aspects of MEC, including security, are hot topics in the research community. This paper focuses on mitigation techniques against Distributed Denial of Service (DDoS) attacks in the context of 5G MEC. One solution against DDoS is to detect network flow anomalies, this being the preferred approach in loaded networks compared to a deep inspection for each packet \cite{flow-detection-survey-and-tutorial}. However, flow inspection does not completely replace deep packet inspection, which performs a more fine-grained detection, so both techniques are useful to mitigate attacks. This paper combines the two techniques and brings the security solutions at the architectural level to mitigate DDoS attacks, being an improvement of the solutions proposed in \cite{dynamic-management}. Starting from the dynamic management approach for the virtualized environment on the MEC hosts developed in the initial study, we propose new solutions on the architecture as well as the management of the virtualized environment to minimize the risk of affecting legitimate users. We thus use two separate virtual machines: one to provide services to legitimate users and one to isolate suspicious traffic for further deep packet inspection. We suggest separating the legitimate traffic from the one detected by the flow collector as an anomaly to protect the first one in case of an attack. If the virtual machine that isolates the anomalies is affected by an attack, it can be reinstated and reconfigured by the involvement of the MEC management entities.  


The paper is organized as follows. Section \ref{sec_background} introduces the necessary background, describing the MEC architecture and reviewing the existing work regarding the detection of DDoS attacks based on the network flow inspection and packet inspection, also offering a refreshing motivation for the need for each approach. Section \ref{sec_solutions} presents the concerns that appear in 5G MEC regarding the detection of DDoS attacks and our contributions, highlighting the improvements over the work done in \cite{dynamic-management}. Section \ref{sec_orchestration} presents the orchestration process at the architectural level, illustrating the role of each MEC entity and the steps followed for the proposed solution. Finally, Section \ref{sec_conclusions} concludes.


\section{Background}
\label{sec_background}

\subsection{MEC architecture}

\begin{figure}[t!]
\begin{center}
\includegraphics[width=0.7\textwidth]{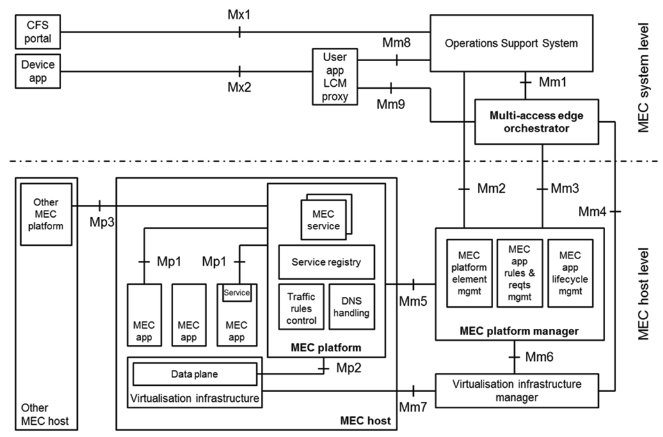}
\caption{MEC architecture \cite{ETSI:MEC003}}
\label{mec-arh-img}
\end{center}
\end{figure}

Figure \ref{mec-arh-img} illustrates the MEC architecture, as introduced by the ETSI specifications \cite{ETSI:MEC003}. Next, we will restrict our presentation to the architectural elements necessary to understand our work. More information on the MEC architecture can be found in \cite{ETSI:MEC003}.

The MEC architecture is split into the \textit{MEC host level} and the \textit{MEC system level}. At the MEC host level, a MEC Host (MEH) implements the MEC applications that run on top of a Virtualized Infrastructure (VI) and are interacting with the MEC Platform (MEP) that offers the necessary environment for running these applications. The MEC applications and the MEP are the ones to offer/consume, respectively host MEC services.

The MEC management resides both at the MEC host level and the MEC system level. At the host level, the MEC Platform Manager (MEPM) and the Virtualization Infrastructure Manager (VIM) are in charge of the functionality of a given MEC host, including the applications that it runs. More precisely, the VIM prepares the virtualization infrastructure and manages the allocation of the resources, while the MEPM manages the life cycle of the applications and further interacts with the MEC Orchestrator (MEO) situated at the system level management. At the system level management, the MEO maintains an overview of the complete MEC system, including overall available resources, services, and it is in charge of triggering the instantiation of the applications (on selected MEC hosts) based on the existing necessities, constraints, and resources. 

\subsection{Network flow analysis and deep packet inspection}


MEC allows a large number of connected devices and, therefore, a high volume of data, while maintaining low-latency. Hence, security solutions must admit the collection, processing, and analysis of a considerable volume of data, which requires the transition from classic \textit{packet inspection} to anomaly detection methods based on \textit{network flow}~\cite{flow-detection-survey-and-tutorial}.


A network flow is defined as a collection of packets that pass through an observation point in the network for a certain period and have in common certain features called \textit{flow keys} \cite{flow-definition}. Flow keys are features generated by using a special feature commonly found in routers, e.g., \textit{NetFlow} \cite{netflow}. These keys can be: IP source address, IP destination address, source port, destination port, protocol, total packet length, but can also be processed and extended to other properties~\cite{flow-definition,flow-agregare}. After the network elements select subsets of packets of interest, a stream of reports on these packets is exported to an external flow collector.


According to \cite{flow-detection-survey-and-tutorial}, the ratio between packets exported as a network flow and whole packets averages $0.1\%$, and the relative size in bytes averages $0.2\%$. Therefore, an approach for detecting network anomalies based on flow analysis is a suitable choice for high-speed networks \cite{dynamic-management,flow-agregare,flow-detection-survey-and-tutorial}. There are also well-organized DDoS attacks that could lead to considering all packets as a network flow and, consequently, a dramatic increase in the data sent for analysis. However, the amount of exported data from these packets is reduced compared to complete packets. Moreover, there are aggregation techniques to combat such events \cite{flow-agregare}.



On the other hand, \textit{deep packet inspection} examines the content of the packet entirely as they pass through the observation point. 
Detection based on network flow analysis does not replace deep packet inspection but offers a possibility of detection when packet analysis is not feasible (e.g., because of too much traffic or time constraints) \cite{flow-detection-survey-and-tutorial}. Therefore, even if the detection accuracy for flow analysis is lower than for packet inspection, the information in the network flow is sufficient to identify patterns in the communication and often leads to the recognition of attacks \cite{flow-detection-survey-and-tutorial}. The approach chosen in this paper respects these aspects and considers a two-phase detection initiated with a network flow analysis and continued with a deep packet inspection if needed.


\subsection{Anomaly detection system in 5G MEC}

\begin{center}
\begin{figure}[b!]
\begin{center}
\includegraphics[width=0.7\textwidth]{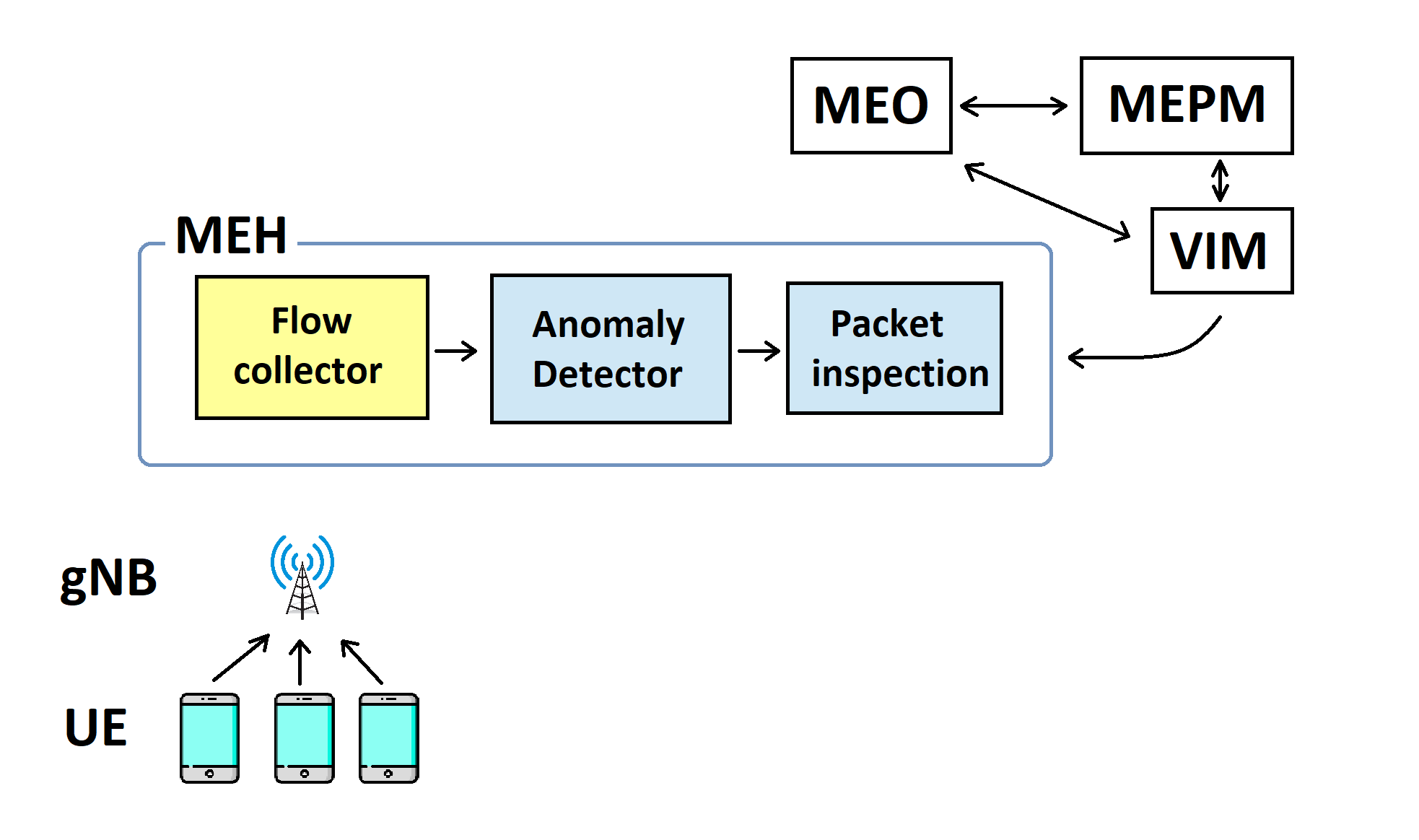}
\caption{Anomaly detection system in 5G MEC}
\label{detectie-ddos-arhitectura-img}
\end{center}
\end{figure}
\end{center}

In consequence, an efficient anomaly detection system in 5G MEC makes use of: (1) a flow collector, (2) an anomaly detector that interprets the flow and decides on the necessity of further deeper inspection, and (3) a deep packet inspection. Our proposed solution comes as an improvement of the solution in~\cite{dynamic-management} to minimize the impact to the legitimate MEC applications and services available on the hosts in case of a DDoS attack.




Figure \ref{detectie-ddos-arhitectura-img} illustrates the anomaly detection system. Our improvement is detailed in Section \ref{sec_solutions}. User Equipment (UE) are devices used by the end-users to perform their tasks (e.g., consume MEC services), accessing the 5G network via a base station called \textit{gNB}. The flow collector is responsible for receiving and storing the network flow, from which it extracts its features in a manner suitable for the anomaly detector that interprets it. The anomaly detector based on network flow analysis focuses on rapid anomaly detection by examining flow features using artificial intelligence solutions. In the event of symptoms corresponding to an attack, a deep packet inspection will be performed. 

If the flow-level detection reveals symptoms of an anomaly, the MEO shall be informed. It will communicate with the MEPM and the VIM to act on the traffic and the rules of the application, respectively, to isolate the detected traffic on another virtual machine for a deep packet inspection. Subsequently, based on the obtained results, further decisions can be made. The isolation of the detected traffic on a separate virtual machine is our proposal to reduce the risk of affecting legitimate MEC applications and services.




\bigskip
\section{Improved solutions}
\label{sec_solutions}

\subsection{Concernes and solutions}
\label{subsec_concerns}

\begin{figure}[t!]
\begin{center}
\includegraphics[width=1.02\textwidth]{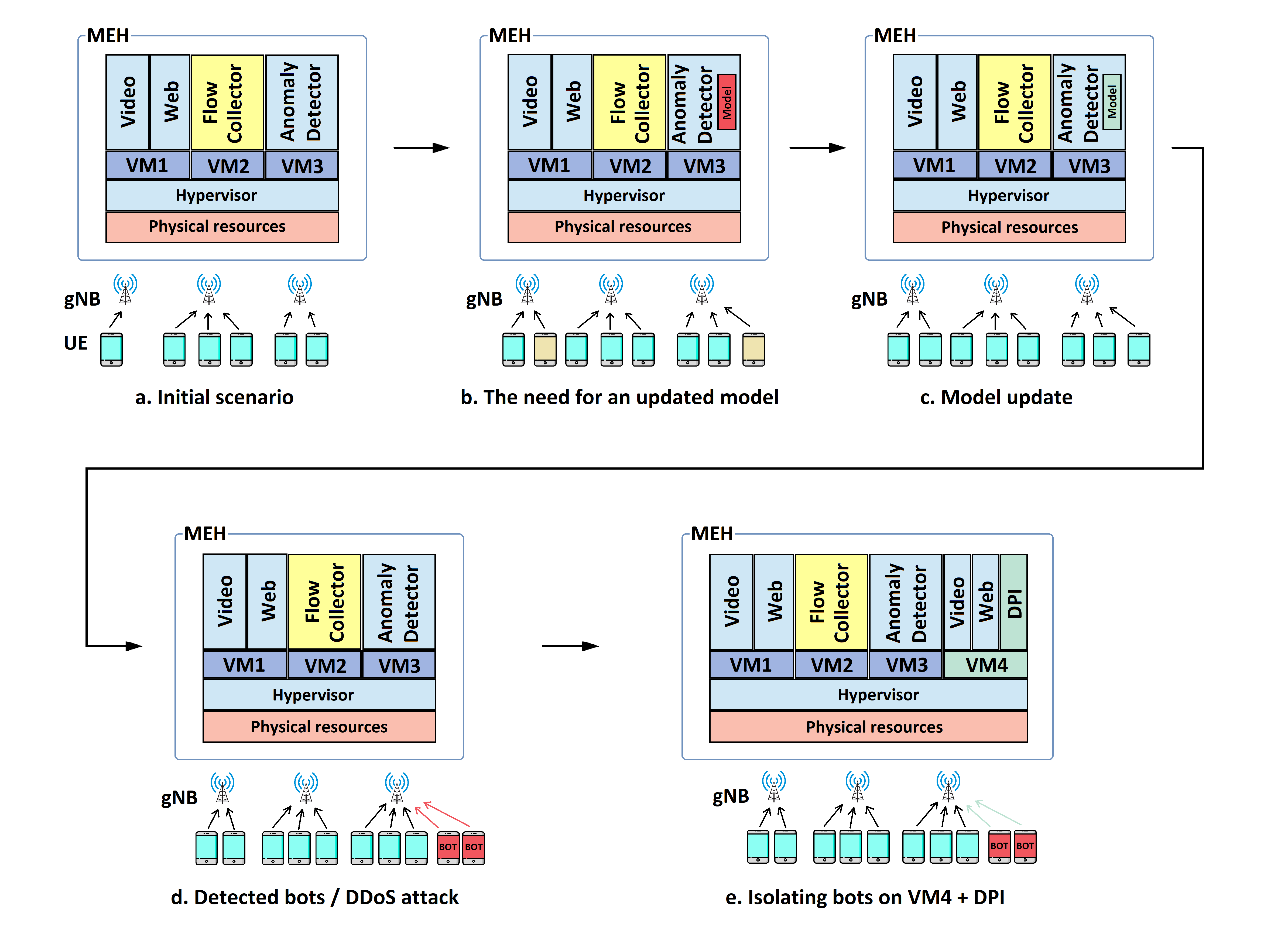}
\caption{Concerns and management solutions (adapted from \cite{dynamic-management} according to the proposed improvements). Red marks the concerns that may arise, respectively, green marks the solutions.}
\label{management-dinamic-img}
\end{center}
\end{figure}

Figure \ref{management-dinamic-img} illustrates the general concerns related to our problem and proposals for solving them in the MEC architecture. It is adapted from  \cite{dynamic-management}, according to the improvements we propose, which are now explained in detail.

Figure \ref{management-dinamic-img}.a shows the initial scenario, in which the users connect to a MEC host to receive services such as internet browsing, email, video/audio streaming services. These services are provided via virtual machines. In this scenario, there is a configuration of three virtual machines with the following functions: VM1 provides the user services, VM2 collects the network flow, and VM3 analyzes the anomalies by inspecting the flow and allows their detection using artificial intelligence techniques. The initial scenario is identical to the one presented in the original paper \cite{dynamic-management}.


Figure \ref{management-dinamic-img}.b highlights the attachment of new devices to the network and the appearance of \textbf{the first concern}, which implies the need to automatically update the anomaly detection module based on the network flow inspection. A key aspect in artificial intelligence techniques is the periodic retraining of models using other data sets to allow continuous learning that helps detect new anomalies \cite{retrain,dynamic-management}. The training process is expensive and requires resources and a large amount of data to obtain favorable results. Therefore, the continuous training of the model takes place on external devices. Thus, a \textbf{solution} is to update the detection module in real-time, by updating the artificial intelligence model and reconfiguring it. The image illustrates the concern with the color red that highlights the old model that needs improvement. 


Figure~\ref{management-dinamic-img}.c marks the solution of launching an updated model on the anomaly detection module, using the color green. This concern, respectively solution, corresponds to the second concern in~\cite{dynamic-management}. We note that we ignore certain concerns present in the original paper because they are redundant for our study.


Figure \ref{management-dinamic-img}.d shows new suspicious devices that have been detected by the anomaly detection module that inspects the network flow. These can be, for example, bots controlled via a Command and Control (C\&C) channel or devices owned by the attacker. In this context, these anomalies are analyzed in detail to conclude if they are indeed a concern and make further decisions. As already mentioned, deep packet inspection is not feasible for all the traffic in 5G MEC because of the considerable amount of packets circulating in the network and the delay it would introduce. Thus appears \textbf{the second concern}, which is to establish anomalies and act to mitigate attacks. 

Figure \ref{management-dinamic-img}.e illustrates the \textbf{proposed solution}. This consists in informing the management entities at the moment of the detection of the possible anomalies so that another virtual machine can be prepared. This virtual machine (VM4) aims to provide services and applications for suspicious devices while performing deep packet inspection. Thus, the traffic of suspicious devices is redirected to this separate machine to be inspected in detail. By isolating the suspicious traffic, we avoid an attack on VM1, which provides services to all legitimate users. The resources of VM4 (in terms of computation power, storage, etc.) are managed dynamically by the VIM depending on the necessities (e.g., the amount of traffic redirected, the number of quarantined devices). The need for a separate virtual machine to provide services is due to the deep packet analysis of packets. This inspection is not possible on VM3 because there is no complete packet information there. Thus, this solution is an improvement of \cite{dynamic-management}, which performed the detailed inspection directly on VM1. The tools that perform deep packet inspection conclude if it is an attack, and the MEPM is informed to make further decisions (e.g., to restrict the devices controlled by the attacker).

\subsection{Architectural proposals}
\label{subsec_arh_prop}

At the architectural level, we mention two possible constructions:
\begin{enumerate}
    \item VM4 is turned on when needed (when anomalies are detected). Subsequently, it can be reconfigured and can dynamically receive resources at the instruction of the management entities. If the anomalies are all treated, there is no need for quarantined services, and there is no more traffic to be inspected, then VM4 can be stooped to free resources.
    \item VM4 runs permanently, and it is dynamically updated at the instructions of the management entities through reconfiguration and reallocation of resources. 
\end{enumerate}

In both cases, the allocation of the resources is performed by the hypervisor that runs on the MEC host and executes the VIM instructions (see Figure \ref{management-dinamic-img}).

\begin{figure}[t!]
\begin{center}
\includegraphics[width=0.8\textwidth]{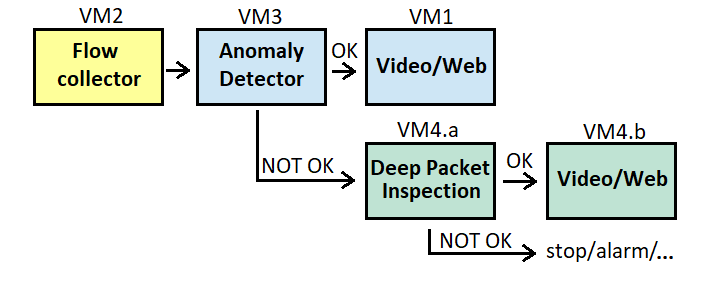}
\caption{Solution scenario}
\label{flow-idee-img}
\end{center}
\end{figure}

A second improvement consists in a separation of the deep packet inspection and the quarantined services by running them on two virtual machines, as illustrated in Figure \ref{flow-idee-img}. This solution provides stability to thorough traffic inspection tools by isolating them on a separate machine. In this scenario, the flow collector sends the flow for analysis to the anomaly detector. If the detector does not find anomalies, the corresponding traffic is considered legitimate, and the services are offered on VM1. Otherwise, the traffic is routed on VM4.a for a deep packet inspection. Following the inspection performed on VM4.a, either the services are quarantined (at least for some time) on a virtual machine VM4.b or the services are dropped, and an alarm is raised. This process is constantly monitored by the management entities to dynamically provide resources, respectively, to reinstate the machines VM4.a and VM4.b in case of damage. Traffic originating from devices that pass VM4.a are not sent back to VM1 because they were once detected as anomalies. In particular, these can be bots that have not yet launched an attack.

\section{The orchestration process}
\label{sec_orchestration}

In this section, we explain the process (present at the MEC architectural level) that allows starting and reconfiguration of virtual machines in case of attacks. The involvement of management entities is necessary: this allows a dynamic reallocation of resources and reinstatement of virtual machines as needed, offering continuous availability in case of an attack. This process follows the proposed solution in the case of anomaly detection, presented in Section \ref{subsec_concerns} but can be easily extended to accommodate the splitting of VM4 in VM4.a and VM4.b as explained in Section \ref{subsec_arh_prop}. Again, the process corresponds to a similar process presented in the original paper \cite{dynamic-management} but incorporates the proposed improvements.



\begin{figure}[t!]
\begin{center}
\includegraphics[width=1\textwidth]{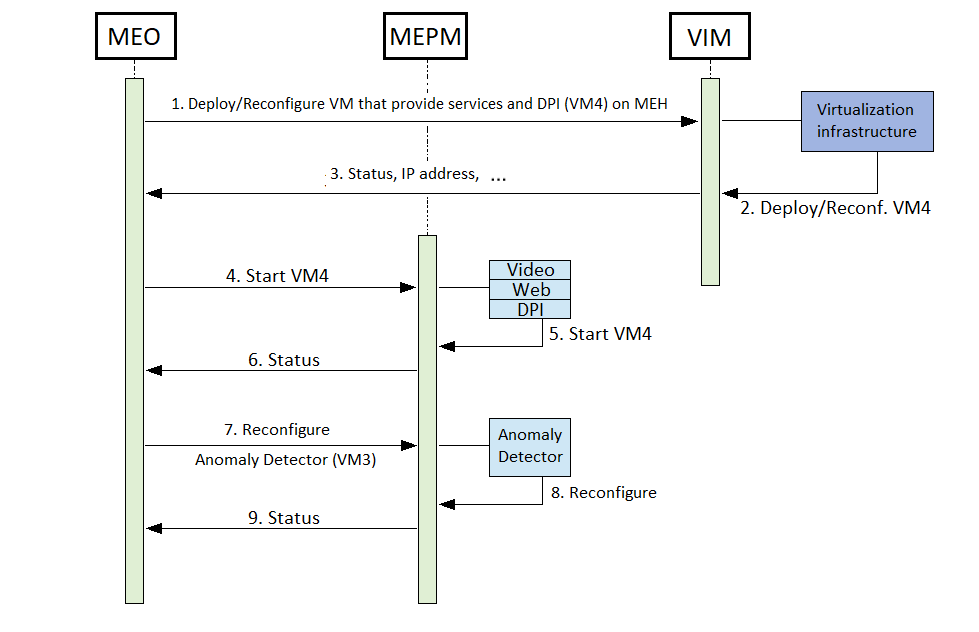}
\caption{Architectural diagram (adapted from \cite{dynamic-management} according to the proposed improvements)}
\label{arhitectura-propusa-img}
\end{center}
\end{figure}

Figure \ref{arhitectura-propusa-img} illustrates the steps followed by the 5G MEC architectural elements in case that the anomaly detection system indicates an attack based on the network flow analysis. The orchestrator (MEO) is informed whenever an anomaly is detected. The MEO interacts with the VIM to instantiate or reconfigure the virtual machine that provides quarantined services and performs the detailed inspection (VM4) on the host in question (MEH) (step 1). The VIM checks if the available physical resources (e.g., memory, processing) are sufficient to instantiate or reconfigure VM4; if so, it applies the changes to the VI (step 2). Furthermore, the VIM informs the MEO about the current status of the machine and provides specific related information (step 3). Once the MEO knows that the machine is turned on and running, respectively its resources are compliant, it sends the machine information to the MEPM to start the services and the deep packet inspection tools (step 4). The MEPM starts or uses the detailed inspection services and tools based on the machine information provided by the MEO (step 5). If the services and DPI already exist on this machine, it is no longer necessary to restart them but only to use them. This holds when the machine is reconfigured.  The MEPM informs the MEO about the status of its actions (step 6). If the status is favorable, the MEO communicates with the MEPM to reconfigure the anomaly detector (step 7). The reconfiguration consists in informing the anomaly detector about the fact that the redirection of the detected traffic as an anomaly on VM4 can start because all the measures have been taken in this regard (step 8). Finally, the MEPM sends to the MEO the reconfiguration status (step 9).

\section{Conclusions}
\label{sec_conclusions}

The study of the MEC is constantly evolving, being an active field of research given the promised benefits for 5G. This paper focuses on security solutions that can be adopted in the 5G MEC architecture to mitigate DDoS attacks. Given the large number of devices and the high volume of traffic in 5G, network flow analysis is proved useful in DDoS mitigation. Starting from existing work [5] that benefits of the virtualized environment, the orchestration, and the addition of artificial intelligence techniques, we offer improved solutions.

More specifically, we proposed two methods to increase protection on services provided to legitimate users. The first proposed method involves the use of a separate virtual machine (VM4) that provides services for the traffic detected as an anomaly, respectively to inspect this traffic in detail and make decisions. The second method comes as an improvement over the previous one, separating the two tasks of the VM4 on two separate virtual machines. This results in VM4.a, which thoroughly inspects traffic, and VM4.b, which provides services for traffic passing the deep packet inspection. Both our methods increase protection against DDoS, increasing the protection of the legitimate services in case of attacks. 

\section*{Acknowledgements}
This work was partially supported by the Norwegian Research Council through the 5G-MODaNeI project (no. 308909).

\bibliographystyle{splncs04}
\bibliography{samplepaper}

\begin{thebibliography}{1}
\providecommand{\url}[1]{\texttt{#1}}
\providecommand{\urlprefix}{URL }
\providecommand{\doi}[1]{https://doi.org/#1}

\bibitem{netflow}
Cisco: {Cisco IOS NetFlow},
  \url{https://www.cisco.com/c/en/us/products/ios-nx-os-software/ios-netflow/index.html},
  {Last accessed: October 2021}

\bibitem{ETSI:MEC003}
ETSI: {GS MEC 003 V2.2.1: Multi-access Edge Computing (MEC); Framework and
  Reference Architecture} (Dec 2020)

\bibitem{flow-definition}
IETF: {RFC 7011 - Specification of the IP Flow Information Export (IPFIX)
  Protocol for the Exchange of Flow Information}

\bibitem{mec2018}
Kekki, S., Featherstone, W., Fang, Y., Kuure, P., Li, A., Ranjan, A.,
  Purkayastha, D., Jiangping, F., Frydman, D., Verin, G., et~al.: {MEC in 5G
  networks}. ETSI white paper  \textbf{28},  1--28 (2018)

\bibitem{dynamic-management}
Maim{\'o}, L.F., Celdr{\'a}n, A.H., P{\'e}rez, M.G., Clemente, F.J.G.,
  P{\'e}rez, G.M.: Dynamic management of a deep learning-based anomaly
  detection system for 5g networks. Journal of Ambient Intelligence and
  Humanized Computing  \textbf{10}(8),  3083--3097 (2019)

\bibitem{mec2014}
{Patel}, M., {Naughton}, B., {Chan}, C., {Sprecher}, N., {Abeta}, S., {Neal},
  A.: {Mobile-Edge Computing – Introductory Technical White Paper}. White
  Paper, Mobile-edge Computing (MEC) Industry Initiative  (Sep 2014)

\bibitem{flow-agregare}
Song, S., Chen, Z.: Adaptive network flow clustering. In: 2007 IEEE
  International Conference on Networking, Sensing and Control. pp. 596--601.
  IEEE (2007)

\bibitem{flow-detection-survey-and-tutorial}
Sperotto, A., Schaffrath, G., Sadre, R., Morariu, C., Pras, A., Stiller, B.: An
  overview of ip flow-based intrusion detection. IEEE communications surveys \&
  tutorials  \textbf{12}(3),  343--356 (2010)

\bibitem{retrain}
Wu, Y., Dobriban, E., Davidson, S.: {DeltaGrad: Rapid retraining of machine
  learning models}. In: International Conference on Machine Learning. pp.
  10355--10366. PMLR (2020)

\end{thebibliography}
\end{document}